\documentclass[reprint,aps,prb,floatfix]{revtex4-2}
\usepackage{bm}
\usepackage{amssymb}
\usepackage{amsmath}
\usepackage{graphicx}
\usepackage{rotating}
\usepackage{epsfig}
\usepackage{psfrag}
\usepackage{amsmath}
\usepackage{subfigure}
\usepackage{braket}
\usepackage{tikz}
\usepackage{stmaryrd}
\usepackage{wasysym}

\newcommand{\bea}{\begin{eqnarray}}
\newcommand{\eea}{\end{eqnarray}}
\newcommand{\bpm}{\begin{pmatrix}}
\newcommand{\epm}{\end{pmatrix}}

\usepackage[unicode=true,colorlinks=true,pdfpagemode=UseOutlines,pdfstartview=Fith]{hyperref}
\hypersetup{linkcolor=blue,citecolor=blue,urlcolor=blue}

\begin{document}
\title{Topological and magnetic phase transitions in the bilayer Kitaev-Ising model}

\author{Aayush Vijayvargia$^1$, Urban F. P. Seifert$^2$, Onur Erten$^1$}
\affiliation{$^1$Department of Physics, Arizona State University, Tempe, AZ 85287, USA \\ $^2$Kavli Institute for Theoretical Physics, University of California, Santa Barbara, CA 93106, USA }

\begin{abstract}
 
We investigate the phase diagram of a bilayer Kitaev honeycomb model with Ising interlayer interactions, deriving effective models via perturbation theory and performing Majorana mean-field theory calculations. We show that a diverse array of magnetic and topological phase transitions occur, depending on the direction of the interlayer Ising interaction and the relative sign of Kitaev interactions. When two layers have the same sign of the Kitaev interaction, a first-order transition from a Kitaev spin liquid to a magnetically ordered state takes place. The magnetic order points along the Ising axis and it is (anti)ferromagnetic for (anti)ferromagnetic Kitaev interactions. However, when two layers have opposite sign of the Kitaev interaction, we observe a notable weakening of magnetic ordering tendencies and the Kitaev spin liquid survives up to a remarkably larger interlayer exchange. Our mean-field analysis suggests the emergence of an intermediate gapped $\mathbb{Z}_2$ spin liquid state, which eventually becomes unstable upon vison condensation. The confined phase is described by a highly frustrated $120^\circ$ compass model. We furthermore use perturbation theory to study the model with the Ising axis pointing along $\hat{z}$-axis or lying in the $xy$-plane. In both cases, our analysis reveals the formation of 1D Ising chains, which remain decoupled in perturbation theory, resulting in a subextensive ground-state degeneracy. Our results highlight the interplay between topological order and magnetic ordering tendencies in bilayer quantum spin liquids.
\end{abstract}
\maketitle

\section{Introduction}
Quantum spin liquids (QSLs) are a unique class of phases in quantum magnets that are not uniquely characterized by local order parameters \cite{Broholm_Science2020, Balents_Nature2010, Savary_RepProgPhys2016,moessner_moore_2021}, but instead exhibit long-range entanglement, fractionalization and emergent gauge fields  \cite{Zhou_RMP2017, Knolle_AnnRevCondMatPhys2019, Wen_RMP2017}, which are understood to be stabilized by strong quantum fluctuations.
Since the first proposal for a QSL by Anderson \cite{Anderson_1973} in 1973, there has been remarkable progress in the identification of both theoretical models that may exhibit QSL ground states and the discovery of candidate materials that exhibit experimental signatures which might be compatible with QSL behaviour.
In this regard, the Kitaev model on a honeycomb lattice \cite{Kitaev_AnnPhys2006} plays an exceptional role as a spin model for a QSL that \emph{both} can be solved exactly and may be (approximately) realized in materials, most prominently $\alpha$-RuCl$_3$ \cite{Trebst_20221,Jackeli_2009}.

Further, in recent years, remarkable experimental progress and theoretical analysis has made evident that bilayers and moir\'e superlattices of 2D (van der Waals) materials represent new, adjustable quantum platforms for realizing a myriad of novel phases \cite{Cao2018, Devakul_NatComm2021, Zhao2021}.
While bilayers of electronic materials have widely been explored, investigations of bilayers of frustrated quantum magnets and magnetic moir\'e superlattices are still in their early stages \cite{Hejazi_PNAS2020, Hejazi_PRB2021, Akram_PRB2021, Akram_NanoLett2021, Das_arXiv2023, Xu_NatNano2021, Song_Science2021, xie2022, Akram_arXiv2023_2}.
\begin{figure}[t]
    \centering
    \includegraphics[width=1\linewidth]{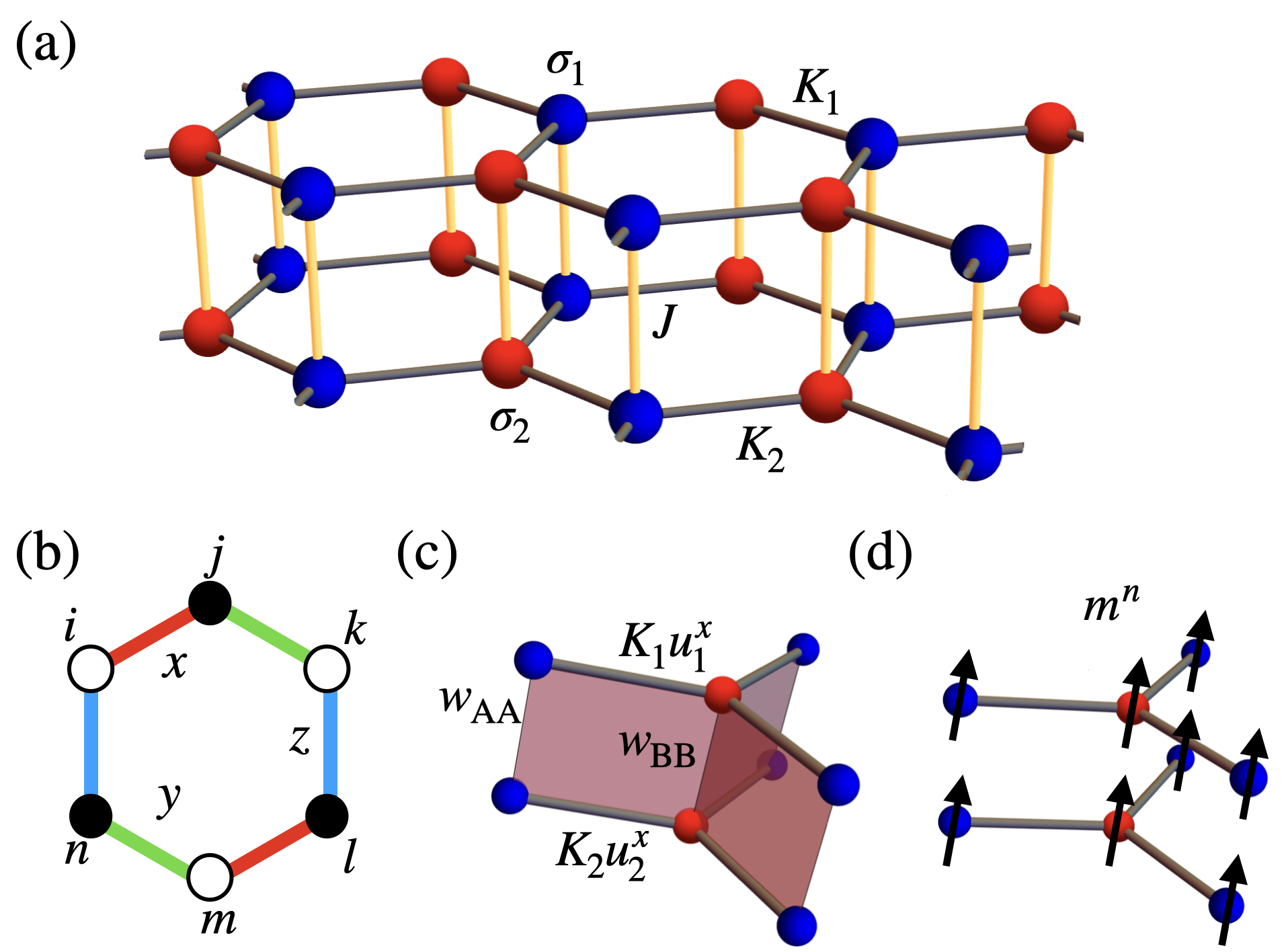}
    \caption{(a) Illustration of the bilayer Kitaev-Ising model. $K$ and $J$ are intralayer Kitaev and interlayer Ising interaction exchange terms, respectively. (b) The bond-dependent interactions of the Kitaev model: red, green and blue bonds represent the $x$, $y$ and $z$ bonds respectively. We observe two types of phase transitions: (c) topological phase transitions where the interlayer plaquettes acquire finite expectation value that gap the spectrum, and (d) magnetic order induced by the interlayer interaction or an external magnetic field.}
    \label{fig:1}
\end{figure}
Considering bilayers of the Kitaev's honeycomb spin liquid \cite{Kitaev_AnnPhys2006}, we note that generic interlayer interactions spoil the integrability of the Kitaev model in each layer \cite{Seifert_PRb2018,May-Mann_PRB2020}, and the resulting model is no longer exactly solvable.
Instead, one can turn to perturbative expansions, starting in solvable limits, perform mean-field treatments \cite{Seifert_PRb2018, Haskell_PRB2022} or use numerical methods such as exact diagonalization \cite{Tomishige_PRB2018, Tomishige_PRB2019} to estimate its phase diagram.
In contrast, $\Gamma$-matrix generalizations of the Kitaev model (with larger local Hilbert spaces) \cite{Nica_npjQM2023, Vijayvargia_PRR2023, Keskiner_PRB2023, Akram_arXiv2023} allow for interlayer exchange terms that commute with the intralayer fluxes, making controlled calculations feasible.
Yet, the lack of candidate materials for these models is a significant challenge.
It is worth noting that prior research has predominantly focused on bilayer Kitaev models with Heisenberg interlayer interactions, which stabilize a trivial quantum paramagnet at for large interlayer interactions, consisting of interlayer singlets  \cite{Seifert_PRb2018, Haskell_PRB2022, Tomishige_PRB2018, Tomishige_PRB2019}.

Instead, in this article, we focus on the $S=1/2$ bilayer Kitaev model with \emph{Ising} interlayer interactions.
Unlike a Heisenberg interlayer interaction, this interaction  retains a residual degree of freedom in the limit of large interlayer exchange couplings.
This opens up the possibility for non-trivial phases in this limit, in particular one may wonder if topologically ordered states or magnetic phases are realized.
In particular, the coexistence of topological and magnetic order could give rise to a spontaneously-generated chiral spin liquid.
In principle, there exists an arbitrariness to fixing the spin-space axis of the Ising interlayer. We note that varying this axis and different choices for the relative sign of the Kitaev couplings add layers of complexity to our investigation, providing the means for exploration of rich phase diagrams and emergent phenomena. 

To construct the phase diagram of the model, we first focus on deriving effective Hamiltonians in the limit of large interlayer exchange interactions.
This allows us to determine the ground state in this limit, such as ferromagnetic (FM) or antiferromagnetic (AFM) order.
Equipped with these controlled insights, we perform Majorana mean-field theory to determine the phase at weak and intermediate interlayer exchange, where we use magnetically ordered states as variational ansatze. We emphasize that by construction, the mean-field theory exactly reproduces the $T=0$ ground state of the Kitaev model (i.e. in the lowest flux sector) and is thus controlled in \emph{both} limits of vanishing and strong interlayer interactions.

Our main results can be summarized as follows: (i) when the Ising interaction points along $(n^x,n^y,n^z)$ with all $n^\alpha \neq 0$, and both layers have the same Kitaev interaction strength $(K_1=K_2)$, there is a first order transition from a $\mathbb{Z}_2 \times \mathbb{Z}_2 $ spin liquid state to a FM or AFM state, depending on the sign of the Kitaev interaction.
(ii) For $K_1=-K_2$, the magnetic order is suppressed and the spin liquid phase is sustained for fairly large interlayer couplings.
Beyond a critical $J/|K|$, within mean-field theory we find that a gapped bilayer $\mathbb{Z}_2$ spin liquid emerges, locking the gauge structure of the two layers.
This phase then undergoes a confinement-deconfinement transition for larger $J/|K|$. Perturbatively, we show that the large interlayer coupling limit of the confined phase is determined by the $120^\circ$ compass model for the effective degrees of freedom.
(iii) In cases when $\mathbf{n}$ is along a Cartesian axis such as the $\hat{z}$ direction, or perpendicular to it (i.e. $\mathbf{n}$ lies in the $x$-$y$ plane), our perturbative analysis shows the existence of Ising chain with a two-fold ground-state degeneracy per chain.
We find that the splitting of this two-fold ground-state degeneracy by interchain couplings is exponentially small in the length of the chains, and therefore, surprisingly, the system possesses a subextensive ground-state degeneracy (given by effectively decoupled chains) in the thermodynamic limit.

The rest of the paper is organized as follows.
In Sec.~\ref{sec:mod-meth}, we introduce the model and describe our methodology, including the perturbative analysis and Majorana mean field theory.
In Sec.~\ref{sec:results}, we present our results for different parameter regimes. We conclude with a discussion and a summary of our results in Sec.~\ref{sec:conclusion}.

\section{Model \& Methods} \label{sec:mod-meth}

\subsection{Microscopic model} \label{sec:mod-microscopic}
The Kitaev honeycomb model \cite{Kitaev_AnnPhys2006} is a paradigmatic example of a highly frustrated $S=1/2$ model characterized by bond-dependent interactions. Within this model, interactions are defined along three distinct types of bonds originating from each lattice site within the honeycomb lattice, which we denote using the symbols $\alpha=x,\ y,\ z$ as shown in Figs.~\ref{fig:1}(a) and (b).
For the bilayer Kitaev model, we consider an AA stacking configuration, where the A sublattice of the first layer is precisely aligned with the A sublattice of the second layer. The interaction between these two layers is governed by an Ising-type interaction oriented along a specific axis in spin space, characterized by a unit-vector $\mathbf{n}$, which henceforth will be referred to as the Ising axis. The full Hamiltonian is $H=H_K+H_J$, 
\begin{align}
    H_K=&\sum_{\nu, \langle ij\rangle^\alpha} K_\nu S_{\nu i}^\alpha S_{\nu j}^\alpha \label{eq:K} \\
    H_J=&-J\sum_i (\mathbf{n}\cdot\mathbf{S}_{1i})(\mathbf{n}\cdot\mathbf{S}_{2i}),\label{eq:J}
\end{align}
where $\nu=1,2$ is the layer index and $\mathbf{n}$ is a vector on the unit sphere. Before delving into the bilayer model, we first briefly review the solution of the single layer Kitaev model following Ref.~\cite{Kitaev_AnnPhys2006}.
The key observation which leads to the exact solvability is based on the plaquette operators $W_p= \sigma^x_i\sigma^y_j\sigma^z_k\sigma^x_l\sigma^y_m\sigma^z_n$.
These operators commute with the Hamiltonian, and hence the whole Hilbert space can be labelled by the eigenvalues of the plaquette operators. Eq.~\eqref{eq:K} can be solved by representing the spin operators at each site by four Majorana fermions, $2S^\alpha_i=i\chi^\alpha_i\chi^0_i$ where we choose the normalization $(\chi^\mu)^2=1$.
The Majorana representation is overcomplete and the physical Hilbert space can be recovered by projecting states with the operator $P=\prod_i(1+D_{i})/2$ where $D_{ i}=\chi_{ i}^0\chi_{ i}^x\chi_{ i}^y\chi_{ i}^z$, which enforces that the fermion parity on each site is even, $D_i \equiv +1$.
Using the Majorana representation, the Kitaev Hamiltonian can then be written as
\begin{align}
    H_K &= \frac{K}{4} \sum_{\langle i  j \rangle^\alpha} (i\chi^\alpha_i\chi^0_i)(i\chi^\alpha_j\chi^0_j) \nonumber\\
    &\equiv \frac{K}{4} \sum_{\langle i  j \rangle^\alpha} i u_{ij} \chi^0_i \chi^0_j, \label{eq:hk-exact}
\end{align}
where in the second line we have introduced $u^\alpha_{ij} = i \chi^\alpha_i \chi^\alpha_j$. Notably, both $\chi^\mu_j$ and $u_{ij}$ anticommute with the constraint operator $D_i$, and thus it becomes clear that the Majorana fermions carry a $\mathbb{Z}_2$ gauge charge and are coupled to a $\mathbb{Z}_2$ gauge field given by $u_{ij}$, with gauge transformations generated by $D_i$.

The plaquette operators can be represented by the product of the bond operators, $W_p=\prod_p u_{ij} $, corresponding to gauge-invariant Wilson loops in the $\mathbb{Z}_2$ gauge theory. Given that the $W_p$ are conserved, the physical Hilbert space decomposes into distinct sectors labelled by the eigenvalues of $W_p$.
According to Lieb's theorem, the ground state of the Kitaev model lies in the zero-flux sector with all plaquette operators having the eigenvalue $W_p=1$. In this sector, the Majorana fermion dispersion is gapless, and possesses two Majorana-Dirac cones. For a bilayer system with vanishing interlayer couplings $J=0$, there are two copies of gapless $\mathbb{Z}_2$ QSLs, resulting in a $\mathbb{Z}_2 \times \mathbb{Z}_2$ phase.

\subsection{The limit of large interlayer exchange}
In the atomic limit with $K_\nu=0$ and $J\neq 0$, the effective degrees of freedom are determined by the Ising interlayer interaction.
The ground state is a doublet given by $\ket{\uparrow^n_1\uparrow^n_2}$ and $\ket{\downarrow^n_1\downarrow^n_2}$ where $\ket{\uparrow^n}$ and $\ket{\downarrow^n}$ are eigenstates of $(\mathbf{n}\cdot \mathbf{S})$, that is the spin operator aligned to the $\mathbf{n}$ Ising axis.
The excited states also form a doublet, $\ket{\uparrow^n_1\downarrow^n_2}$ and $\ket{\downarrow^n_1\uparrow^n_2}$.
For later convenience, we rotate the axis of quantization of the Pauli matrices such that the rotated $\hat{z}$-axis point along $\mathbf{n}$. To achieve this, we choose the axis of rotation and the angle to be $\mathbf{k}=\mathbf{n}\times \hat{z}$ and $\theta=\cos^{-1}(n^z)$.
Next, we use the operator $\exp(i\theta/2 \mathbf{k}\cdot \boldsymbol{\sigma})$, to rotate each spin matrix at every site along the desired axis using the relation: $
e^{-i\boldsymbol{\sigma}\cdot\hat{\textbf{k}}\theta/2}  \textbf{a}\cdot\boldsymbol{\sigma}   e^{i\boldsymbol{\sigma}\cdot\hat{\textbf{k}}\theta/2}=\left[ 
\hat{\textbf{k}}(\hat{\textbf{k}}\cdot \textbf{a} )+
\cos(\theta)(\textbf{a}-\hat{\textbf{k}}(\hat{\textbf{k}}\cdot \textbf{a}))
+\sin(\theta)\hat{\textbf{k}}\times\textbf{a} 
\right]\cdot\boldsymbol{\sigma}$.
This rotation maps $(\textbf{n}\cdot \boldsymbol{S}) \rightarrow \Tilde{S}^z$, and then the interlayer interaction can be written as $H_J=-J\sum_i\Tilde{S}_{1i}^z\Tilde{S}_{2i}^z$.

In the following, we will derive effective Hamiltonians within the degenerate ground-state manifold spanned by degenerate doublets on each site. To this end, it will be convenient to introduce pseudospin operators for each interlayer pair of sites. These span a full operator basis for each local ground state doublet,
\begin{align}
    \eta^z=&\frac{1}{2}(\Tilde{S}_{1i}^z+\Tilde{S}_{2i}^z) \nonumber\\
     \eta^x_{i}=& \frac{1}{4}(\Tilde{S}_{1i}^x\Tilde{S}_{2i}^x-\Tilde{S}_{1i}^y\Tilde{S}_{2i}^y ) \nonumber\\
    \eta^y_{i}=& \frac{1}{4}(\Tilde{S}_{1i}^x\Tilde{S}_{2i}^y+\Tilde{S}_{1i}^y\Tilde{S}_{2i}^x)
    \label{etadef}
\end{align}
These pseudospin operators satisfy the SU(2) algebra.
Note that $\eta^z$ is a dipolar operator while $\eta^{x}$ and $\eta^y$ are quadrupolar operators \cite{Vijayvargia_PRR2023}.
If the sign of $J$ is flipped from positive to negative, the pseudospin operators need to be redefined as the ground state sector will then be spanned by $\ket{\uparrow^n_1\downarrow^n_2}$ and $\ket{\downarrow^n_1\uparrow^n_2}$. 
The effective Hamiltonian acting on this degenerate subspace, obtained via perturbation theory in the large $J/K|$ limit, can be expressed using these operators \cite{Vijayvargia_PRR2023}.
For instance, the first and second order contribution to the effective Hamiltonian are derived as: 
\begin{align}
    H_{\rm eff}^{(1)}&=P_0H_KP_0 \nonumber \\
    H_{\rm eff}^{(2)}&=P_0H_KSH_KP_0,
\end{align} 
where we use projection operator (in the rotated basis): $P_0= \prod_i (1+4\Tilde{S}_{1i}^z\Tilde{S}_{2i}^z)$ onto the low-energy manifold, and $S=(1-P_0)/(E_0-H_J)$.
We stop at the order of perturbation when the effective Hamiltonian exhibits non-trivial magnetic order. 
If $H_{\rm eff}$ has a simple form (i.e. without frustrated interactions), the ground state in the $J/K \gg 1$ limit can then be readily obtained. 
We will use the thus-obtained magnetically ordered states as ansatze in our Majorana mean-field theory calculations to explore the weak and intermediate $J/K$ regions. 

\subsection{Majorana mean field theory}

In the presence of interlayer interactions, the single-layer Kitaev model as detailed in Sec.~\ref{sec:mod-microscopic} is no longer solvable, as the plaquette operators are no longer conserved, $[H_J,W_p] \neq 0$. To map out phase diagrams, we therefore resort to Majorana mean-field theory (MMFT) for the full bilayer system \cite{Seifert_PRB2018a}.
In the following, we also incorporate an onsite external magnetic field into the Hamiltonian, which will find utility in specific sections of our analysis.

Within MMFT, we do not enforce the constraint $D_i = +1$ for each site (which would require significant numerical effort, e.g.~using Gutzwiller-projected variational Monte Carlo methods), but instead enforce the constraint on average. To this end, we reformulate $D_i=1$ as $i\chi^\alpha\chi^0+ \frac{i}{2}\epsilon^{\alpha \beta \gamma} \chi^\beta\chi^\delta=0$ and subsequently enforce it through the introduction of a Lagrange multiplier, as detailed in Refs. \onlinecite{Seifert_PRB2018a, Yilmaz_PRR2022}.
 
To facilitate the analysis, we employ a mean-field approximation to decouple intralayer Majorana fermion interactions as
$i\chi_{i,A}^\alpha \chi_{i,A}^0i\chi_{j,B}^\alpha\chi_{j,B}^0\approx m^\alpha_A(i\chi_{j,B}^\alpha \chi_{j,B}^0)+m^\alpha_B(i\chi_{i,A}^\alpha \chi_{i,A}^0)-m^\alpha_Am^\alpha_B-u^\alpha(i\chi^0_{i,A}\chi^0_{j,B})-u^0(i\chi^\alpha_{i,A}\chi_{j,B}^\alpha)+u^\alpha u^0$, with the mean-field parameters $u^0=\langle i\chi^0_{i,A}\chi^0_{j,B}\rangle, \  m^\alpha_A=\langle i \chi_{i,A}^\alpha \chi_{i,A}^0 \rangle $ and $m^\alpha_B=\langle i \chi_{j,B}^\alpha \chi_{j,B}^0\rangle$.
The interlayer interaction, $H_J=-\frac{J}{4} \sum_i(i\chi_{1i}^{(\mathbf{n})}\chi^0_{1i})(i\chi_{2i}^{(\mathbf{n})}\chi^0_{2i})$, where $\chi^{(\mathbf{n})}=\sum_\alpha n^\alpha\chi^\alpha$, is decoupled as $i\chi_{1i}^{(\mathbf{n})}\chi^0_{1i}i\chi_{2i}^{(\mathbf{n})}\chi^0_{2i}\approx w_i^0(i\chi_{1i}^{(\mathbf{n})}\chi_{2i}^{(\mathbf{n})})+w_i^{(\mathbf{n})}(i\chi^0_{1i}\chi^0_{2i})-w_i^{(\mathbf{n})}w_i^0-m_{1i}^{(\mathbf{n})}(i\chi_{2i}^{(\mathbf{n})}\chi^0_{2i})-m_{2i}^{(\mathbf{n})}(i\chi_{1i}^{(\mathbf{n})}\chi^0_{1i})+m^{(\mathbf{n})}_{1i}m^{(\mathbf{n})}_{2i}$
with $w_i^0=\langle i\chi^0_{1i}\chi^0_{2i} \rangle$, $w_i^{(\mathbf{n})}=\langle i\chi_{1i}^{(\mathbf{n})}\chi_{2i}^{(\mathbf{n})} \rangle$ denoting mean fields in the Hartree channel, while $m_{1i}^{(\mathbf{n})}=\langle i\chi_{1i}^{(\mathbf{n})}\chi^0_{1i}\rangle $ and $m_{2i}^{(\mathbf{n})}=\langle i\chi_{2i}^{(\mathbf{n})}\chi^0_{2i}\rangle $ is the decoupling in the magnetic channel.
Note that, $m^{(\mathbf{n})}$ is the magnetization along the direction of the $\mathbf{n}$ axis and $m^\alpha$ is the magnetization along $x,~ y,\ z$ axes.
Incorporating all these, we write down the full mean-field Hamiltonian as 
\begin{widetext}
\begin{eqnarray}
 H=&&\sum_{\nu,{i}}\sum_{\alpha-\rm bonds} \frac{1}{2} \left(\frac{K_\nu}{2}m_B-h^\alpha-\lambda^\alpha\right)i\chi^\alpha_{\nu i,A}\chi^0_{\nu i,A}+\frac{1}{2}\left(\frac{K_\nu}{2}m_A-h^\alpha-\lambda^\alpha\right)i\chi^\alpha_{\nu j,B}\chi^0_{\nu j,B} -\frac{K_\nu u^\alpha}{4}\left(i\chi^0_{\nu i,A}\chi^0_{\nu j,B}\right)\nonumber\\
 &&-\frac{K_\nu u^0}{4}\left(i\chi^\alpha_{\nu i,A}\chi_{\nu j,B}^\alpha\right) - \lambda^\alpha\frac{\epsilon^{\alpha\beta\gamma}}{4}\left(i\chi_{\nu i,A}^\beta\chi_{\nu i,A}^\gamma+i\chi_{\nu j,B}^\beta\chi_{\nu j,B}^\gamma\right)\nonumber\\
 &&-\frac{J}{4}\sum_i w_i^0 \left(i\chi_{1i}^{(\mathbf{n})}\chi_{2i}^{(\mathbf{n})}\right)+w_i^{(\mathbf{n})}\left(i\chi^0_{1i}\chi^0_{2i}\right)-m_{1i}^{(\mathbf{n})}\left(i\chi_{2i}^{(\mathbf{n})}\chi^0_{2i}\right)-m_{2i}^{(\mathbf{n})}\left(i\chi_{1i}^{(\mathbf{n})}\chi^0_{1i}\right)+E_\mathrm{const}[m,u,w],
 \label{eqn:Hmf}
\end{eqnarray}
\end{widetext}
where in total 8 mean-field parameters $u,w,m$ and 3 Lagrange multipliers $\lambda^\alpha$ are to be determined self-consistently. $E_\mathrm{const}[m,u,w]$ is a constant term that depends on the mean field parameters.
We use an iterative procedure to solve the mean-field self-consistency equations and determine the Lagrange multipliers, where we diagonalize Eq.~\eqref{eqn:Hmf} on momentum space grids of $4\times 10^4$ points.

As discussed in previous works, the mean-field decoupling of the single-layer Kitaev interaction in Eq.~\eqref{eqn:Hmf} can be seen to exactly reproduce static spin-spin correlations and the spectrum of the itinerant Majorana fermions in the 0-flux ground state sector \cite{Kimchi_PRB2012,Choi_PRB2018}, where intuitively the mean-field parameter $u^\alpha$ can be identified with a (gauge-fixed) configuration of the gauge field $u_{ij}$ in Eq.~\eqref{eq:hk-exact}. 

Next, we present the results obtained using the methods above for various possibilities of $\mathbf{n}$ and the relative sign of the Kitaev interactions in the two layers.

\section{Results} \label{sec:results} 
\subsection{Arbitrary Ising axis with same Kitaev interaction ($n^\alpha \neq 0,\ K_1=K_2$)}
We first consider the case where the Ising interaction has components along all Cartesian coordinates, $\mathbf{n} = (n^x,n^y,n^z)$, with $n^\alpha \neq 0$.
We proceed according to the method described in the previous section, and first derive an effective Hamiltonian for $J/K \gg 1$ via perturbative expansion.
First order perturbation theory leads to
\begin{align}
   H_{\rm eff}^{(1)} =\sum_{\langle ij \rangle^\alpha} U^\alpha \eta_i^z \eta_j^z  \label{eq:Heff1}            
\end{align}
where $U^\alpha= (K_1+K_2)(n^\alpha)^2/2$.
For the isotropic direction, $\mathbf{n}=(1,1,1)/\sqrt{3}$, and $K_1=K_2=K$ we obtain $U^\alpha=K/3$ for all bonds.
Eq.~\eqref{eq:Heff1} suggests the ground state exhibits FM or AFM long range order depending on the sign of K.
It is noteworthy that highly-frustrated Kitaev interactions lead to a simple, non-frustrated effective model in this limit with a straightforward AFM/FM ground state aligned along the Ising axis.

\begin{figure}
    \centering
    \includegraphics[width=1\linewidth]{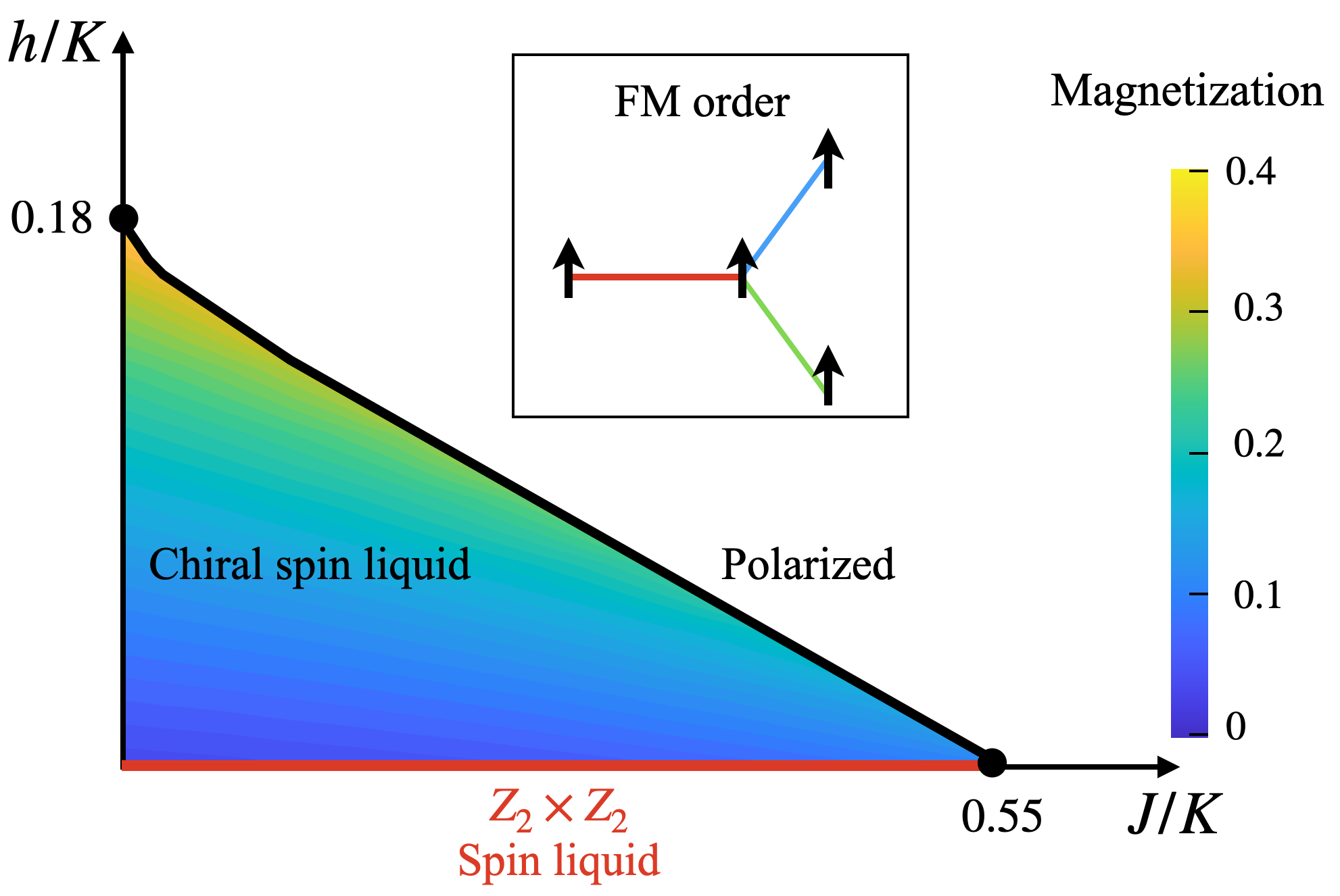}
    \caption{a) Phase diagram for $\mathbf{n}=(1,1,1)/\sqrt{3}$ and $K_1=K_2$. For both FM and AFM Kitaev interactions, the $\mathbb{Z}_2 \times \mathbb{Z}_2$ gapless spin liquid (red line) undergoes a first order transition to a polarized phase as $J_c/K=0.55$. For FM Kitaev interaction, external magnetic field, $h$, along the [111] direction lowers the critical $J$. External magnetic field induces a finite magnetization, which is shown with the colour coding. This phase is a gapped chiral spin liquid.}
    \label{fig:2}
\end{figure}

Next, we perform Majorana mean-field theory calculations to explore the intermediate $J$ region.
We begin with solving the mean field Hamiltonian in Eq.~\eqref{eqn:Hmf}, with no external field.
We find a transition from the $\mathbb{Z}_2 \times \mathbb{Z}_2$ gapless spin liquid, which is characterized by a vanishing magnetisation and no interlayer Hartree channel ($w^{(\mathbf{n})}$ and $w^0$), to a fully polarised state with a uniform magnetization $m^{(\mathbf{n})}=1$.

This holds for both FM or AFM Kitaev interactions. Moreover, this transition depends heavily on the initial conditions, signalling a first-order phase transition. To pinpoint the exact value of $J_c/K$, we compare the energies of the $\mathbb{Z}_2 \times \mathbb{Z}_2$ gapless spin liquid and the fully polarized state and find that the energies intersect at $J_c/K=0.55$ as shown in Fig.~\ref{fig:2}. This demonstrates that, based on our mean-field analysis, we do not expect a phase that simultaneously exhibits local magnetic order and topological order.

Focussing on the case of FM Kitaev interactions, we consider the impact of a magnetic field in the $[111]$ direction. 
In the absence of interlayer interactions ($J=0$), we obtain a chiral spin liquid up to $h_c/K=0.18$, in agreement with Ref. \citenum{Yilmaz_PRR2022}.
With the inclusion of interlayer couplings, $h_c$ diminishes, as expected since the FM interlayer exchange functions similar to magnetic field at mean-field level, leading to a higher effective magnetic field experienced by each layer.
We also observe that if the magnetic channel is artificially turned off, the interlayer Hartree channel acquires a finite expectation value at $J/K=0.9$. 
Given that this value surpasses the critical exchange needed for the fully polarized phase, we can infer that magnetic ordering is preferred compared to the interlayer Hartree channel.

It is important to note that Majorana mean-field calculations on the Kitaev model tend to overestimate the critical values for the destruction of the Kitaev QSL phase, since they ignore the quantum fluctuations due to dynamical visons as excitations of the $\mathbb{Z}_2$ gaiuge field \cite{Tomishige_PRB2019,Trebst_2019,Pollman_2018}. An appropriate treatment is an interesting direction for future research.
Nevertheless, the phase diagrams of mean-field calculations and numerical approaches can be expected to be similar, with renormalized values for the critical coupling constants.

\subsection{Suppressed magnetic ordering for Kitaev interaction with opposite sign ($K_1=-K_2$)} \label{sec:opposite-sign}

Eq.~\eqref{eq:Heff1} implies that the first order correction in the effective Hamiltonian vanishes when $K_1=-K_2$.
Motivated by this observation, we investigate the phase diagram for $K_1=-K_2=K$ and $\mathbf{n}=(1,1,1)/\sqrt{3}$.
Then, second order perturbation theory leads to the following effective spin Hamiltonian in the large $J$ limit,
\begin{align}
    H_{\rm eff}^{(2)}=\frac{2K^2}{|J|}[&\sum_{\langle ij\rangle^z} \eta^x_i\eta^x_j + \sum_{\langle ij\rangle^x} R^{z}_{120}(\eta^x_i)R^{z}_{120}(\eta^x_j)\nonumber \\&+\sum_{\langle ij\rangle^y}  R^{z}_{-120}(\eta^x_i)R^{z}_{-120}(\eta^x_j)]
    \label{eq:H120}
\end{align}
where $R^z_{\theta}(\eta^x)=\exp(i\frac{\theta}{2}\eta^z)\eta^x\exp(-i\frac{\theta}{2}\eta^z)$ is the rotation operation on the pseudospin operators about the $\hat{z}$-axis by $\theta=\pm120^\circ$.
Notably, Eq.~\eqref{eq:H120} is the $120^\circ$ compass model for the $\eta$ degrees of freedom.
It is a highly-frustrated model and its ground state has still not been unambiguously identified.
Candidate orders include valence bond solid, long-range dimer order \cite{Zou_2016,Lou_2015}. 

\begin{figure}
\centering
\includegraphics[width=1\linewidth]{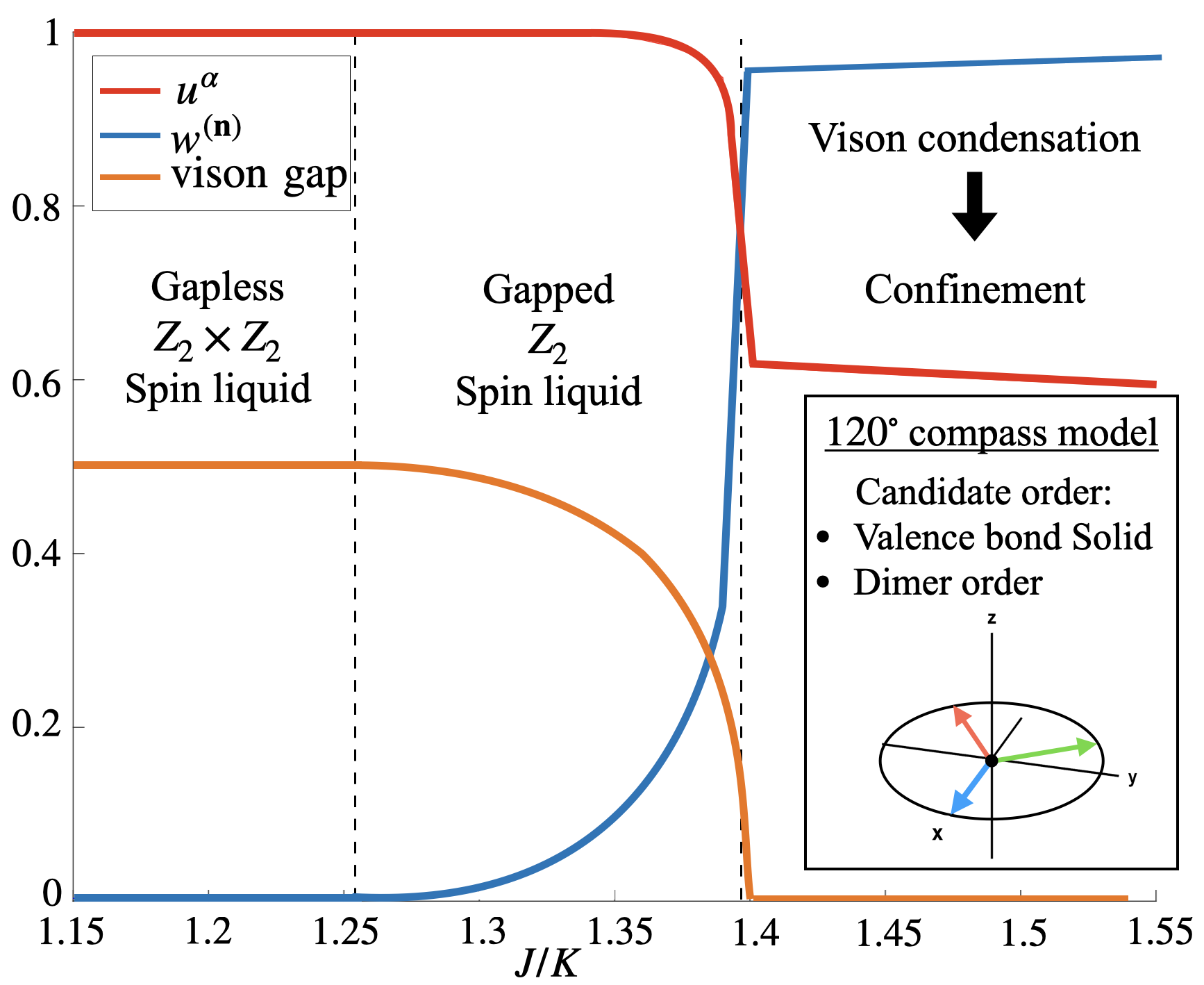}
\caption{ Phase diagram $K_1=-K_2$ and $\mathbf{n}=(1,1,1)/\sqrt{3}$. At $J/K=1.25$, the Hartree order parameter $w^\mu$ acquires a finite expectation value which gaps the spectrum and locks the gauge fields on each layer.
This is followed by a confinement-deconfinement transition via the condensation of visons, which occurs (in a treatment beyond mean-field theory) when the vison gap closes. Here, we take the energy gap of the $\chi^\alpha$-bands as a proxy for the energy cost of a single vison excitation in the full interacting $\mathbb{Z}_2$ gauge theory.
Using perturbation theory, we predict that this phase, at large values of $J/K$, is described by the $120^\circ$ compass model. 
}
    \label{fig:3}
\end{figure}

Since the ground state of the $120^\circ$ compass model is not well-established, a major reason being that the energy differences between the candidate magnetic orders are quite small, we instead use for simplicity FM and AFM (Néel order) mean field ansatze for our mean field theory calculations: $m^{\mathbf{(n)}}_{\nu,A}=m^{\mathbf{(n)}}_{\nu,B}, \text{for FM and} \  m^{\mathbf{(n)}}_{\nu,A}=-m^{\mathbf{(n)}}_{\nu,B}$ for AFM case. We find that these magnetically polarized phases exhibit higher energies compared to the $K_1=K_2$ case, since the energy gain from the Kitaev term on each layer cancels each other due to the opposite sign. 
This allows for the Hartree channel order parameter, $w^\mu$, to attain a finite expectation value prior to magnetic order. Consequently, the interlayer plaquette operator, as shown in Fig. \ref{fig:1}(c), attains a non-zero value, leading to a topologically trivial gapped QSL at $J_c/K=1.25$ as shown in Fig.~\ref{fig:3}. 

We now comment on the interpretation of our results beyond the mean-field treatment of the model.
The mean-field Hamiltonian in Eq.~\eqref{eqn:Hmf} can be understood to constitute a particular gauge-fixed configuration of some (non-integrable) gauge theory.
Equivalence classes of such mean-field ansatze which are equivalent (up to gauge transformations) can be classified with respect to their projective symmetry group (PSG) \cite{Kimchi_PRB2012}.
We refrain from a such a full classification for the bilayer system here.
However, importantly, we note that a finite $w^\mu$ implies that independent gauge transformations on each layer no longer leave the Hamiltonian invariant, only conjoint gauge transformations do.
This reduces the gauge group from $\mathbb{Z}_2\times\mathbb{Z}_2$ to $\mathbb{Z}_2$ \cite{Vijayvargia_PRR2023}.
Moreover, we stress that the operators $W_i^\mu=i\chi^\mu_{1i}\chi^\mu_{2i}$ are in general not gauge-invariant, and thus the fields $w_i^\mu$ can not be used to construct a local Landau-Ginzburg analysis for the transition out of the $\mathbb{Z}_2 \times \mathbb{Z}_2$ spin liquid to the bilayer system with a residual $\mathbb{Z}_2$ gauge group.
Explicitly, gauge transformations induced by the operators $D_{\nu i}$ change the sign of the associated $w_i^\mu$, in addition to the three bond operators, $u_{ij}^\alpha$ emanating from that site.
Consequently, $w_i^\mu$ vanishes for the physical wave function which is symmetrized over all gauge configurations \cite{Kitaev_AnnPhys2006}.
However, it is possible to introduce a gauge-invariant correlator \cite{Nica_npjQM2023,Vijayvargia_PRR2023},
\begin{align}
    \langle C_{ij}^\mu \rangle = \langle W^\mu_i B_{ij}W^\mu_j\rangle
\end{align}
where $B_{ij}= \prod_{\langle i'j'\rangle} \mathrm{sgn}(u^\mu_{1i'j'})\mathrm{sgn}(u^\mu_{2i'j'}) $, is the product of the signs of the $u^\mu_{\nu ij}$ operators that connect the two $W_{i/j}^\mu$ operators.
The value of $B_{ij}$is the same in all gauge choices.
Therefore, it is also finite for the physical wave function.
Finite $w_{i/j}^\mu\neq 0$ implies $\langle C_{ij}^\mu \rangle \neq 0$, signalling a non-local string order parameter. 

For larger values of interlayer exchange, we observe that the energy gap of $\chi^\alpha$ bands vanishes as shown in Fig.~\ref{fig:3}.
These bands are associated with the Majorana fermions of flavor $\alpha$ that are localized on the $\alpha$-bonds in the pure Kitaev limit, which in the exact solution give rise to the $\mathbb{Z}_2$ gauge field (compare also Eq.~\eqref{eq:hk-exact}).
While the vison in Kitaev's exact solution is a \emph{non-local} excitation of the $\mathbb{Z}_2$ gauge field, the delocalization of the $\alpha$-Majoranas (i.e. dispersive bands) can be taken as a proxy for the dynamics of the visons that is induced by breaking integrability, and we therefore (loosely) associate the gap of the $\chi^\alpha$-Majorana fermion dispersion with the gap of dispersing visons in the full (non-integrable) $\mathbb{Z}_2$ gauge theory.
Equipped with this understanding , we suggest that the $\chi^\alpha$-Majoranas becoming gapless can be interpreted as the single-vison gap closing, which allows for the condensation of visons, tantamount to a confinement-deconfinement transition \cite{Sachdev_2011,Senthil_2000}. From our mean-field computations, we find a critical coupling of approximately $J/K\simeq 1.4$.
The resulting state will be accurately described by the $120^\circ$ compass model, as presented in Eq.~\eqref{eq:H120}, for which previous studies have identified non-fractionalized states with magnetic/VBS ordering as possible ground states. 
\subsection{Special cases for the Ising axis}

The first order correction to the effective Hamiltonian in Eq.~\eqref{eq:Heff1} also becomes suppressed if the Ising axis is oriented such that $n^\alpha$ ($\alpha = x,~y,~z$) vanishes for certain bonds.
Unlike the $(K_1=-K_2)$ case in Sec.~\ref{sec:opposite-sign}, where $H_{\rm eff}^{(1)}$ vanishes entirely, orienting the $\mathbf{n}$ such that $n^\alpha=0$ for particular Cartesian axes only suppresses the bonds along the $\alpha$ directions.
To investigate the consequences of these interactions, we consider two cases, where $n^\alpha=0$ for one and two Cartesian axes, respectively, below.

\subsubsection{Effective chain geometry for $\mathbf{n}=(1,1,0)/\sqrt{2}$}

\begin{figure}
    \centering
    \includegraphics[width=1\linewidth]{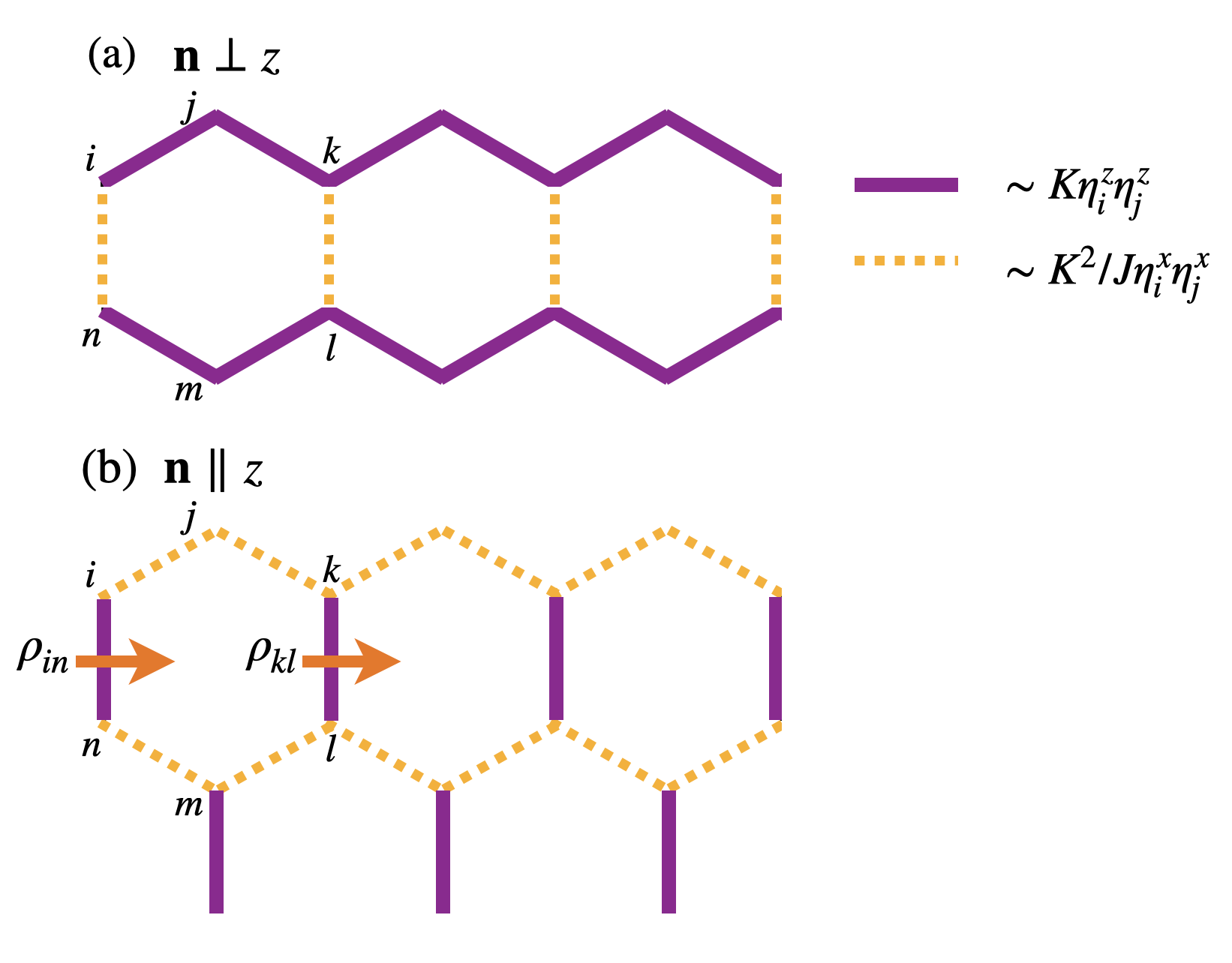}
    \caption{Depiction of the $H_{\rm eff}^{(2)}$ for special cases of the Ising axis. Purple and dashed orange lines represent first order and second order terms in the effective Hamiltonian. (a) For the $\mathbf{n}=[1,1,0]/\sqrt{2}$, The first order correction forms Ising chains along $x$ and $y$. These chains remain decoupled within perturbation theory. (b) For $\mathbf{n}=[0,0,1]$, the first order correction lead to formation of Ising dimers. (c) These dimers couple to form chains in fifth order in perturbation theory. Once again, the chains remain decoupled perturbatively, leading to subextensive degeneracy in both cases.}
    \label{fig4}
\end{figure}

We first consider the case when a single $n^\alpha$ vanishes. We pick $\mathbf{n}=(1,1,0)/\sqrt{2}$, which preserves the symmetry between the $x$ and $y$ bonds, but the first order correction the energy along the $z$ bond vanishes.
We obtain the following effective Hamiltonian up to second order in perturbation expansion, 
\begin{align}
    H_{\rm eff}^{ (1)} &= K\sum_{\langle ij\rangle^{x/y}} \eta_i^z\eta_j^z \nonumber \\ H_{\rm eff}^{ (2)} &=\frac{2K^2}{|J|}\sum_{\langle ij\rangle^{x/y}} \eta_i^x\eta_j^x \hspace{20 pt} ( x/y \ \rm bonds) \label{eq:10}\\
     H_{\rm eff}^{ (2)} &= \frac{2K^2}{|J|}\sum_{\langle ij\rangle^z} \eta_i^x\eta_j^x \hspace{20 pt}\ \ ( z \ \rm bonds)\label{eq:11}
\end{align}
Eq.~\eqref{eq:10} leads to the formation of chains along $x/y$ bonds, coupled along the Ising axis (as depicted in Fig.~\ref{fig4}(a)). This is the largest interaction in the perturbation theory, $\mathcal{O}(K)$, and at this order, each chain exhibits two degenerate ground states.
Meanwhile, at each lattice site, the spins along a chain interact with spins on adjacent chains in the transverse direction in spin space, with a notably diminished interaction strength on the order of $\mathcal{O}(K^2/J)$. Considering the two adjacent Ising chains, a single $H_{\rm eff}^{(2)}$ bond flips two spins and therefore takes the state outside the ground state manifold of Eq.~\eqref{eq:10}.
Consequently, the interchain interactions in $H_{\rm eff}^{(2)}$ do not split the degeneracy between different chains in leading order $K/|J|$.

In order to determine if there are higher-order contributions to $H_\mathrm{eff}$ which lift the degeneracy, we perform exact diagonalization on a 12 site system, which is a single hexagon on both layers. 
We extract the following effective Hamiltonian,
\begin{align}
    H^\mathrm{ED}_{\rm eff}=\sum_{\hexagon} &[c_1K^3/J^2(\eta_i^x\eta_j^x\eta_k^x+\eta_l^x\eta_m^x\eta_n^x) \nonumber\\
    +&c_2K^6/|J|^5(\eta_i^x\eta_j^x\eta_k^x\eta_l^x\eta_m^x\eta_n^x)] \label{HED}
\end{align}
where $c_1\approx 10^{-2}$ and $c_2\approx 10^{-5}$. 
The details of this calculation are given in Appendix \ref{sec:exact-diag}.
While the second term involves interactions between spins on different chains, it flips three bonds on each chain, and therefore takes the chains outside their ground state manifold determined by Eq.~\eqref{eq:10}, similar to $H_{\rm eff}^{(2)}$.

Next, we argue that the degeneracy between distinct chains, determined by $H_\mathrm{eff}^{(1)}$ remains at arbitrarily high-order when including the effects of interchain interactions in $H_\mathrm{eff}^{(2)}$ perturbatively in $K/J \ll 1$. 
To this end, we denote the two degenerate Ising ground states of a chain according to $H_\mathrm{eff}^{(1)}$ as $\ket{\Uparrow(\Downarrow)} = \prod_i \ket{\uparrow(\downarrow)}$. Considering two chains, labelled `t' and `b', interactions lift the four-fold ground state degeneracy if there exists some non-trivial Hamiltonian $\tilde{H}_\mathrm{eff}$ acting on $\ket{\Uparrow_t\Uparrow_b},\dots,\ket{\Downarrow_t\Downarrow_b}$.
We first note that symmetry strongly constrains the form of $\tilde{H}_\mathrm{eff}$: 
Performing a $\pi$-rotation about the $x$-axis of the spins along a given chain, $U=\exp{(-i\frac{\pi}{2}}\sum_{\nu i}\sigma^x_{\nu i})$, flips the spins from $\ket{\Uparrow}\rightarrow\ket{\Downarrow}$ and vice-versa, but commutes both with Eq.~\eqref{eq:11} and any effective Hamiltonian $H^{(n)}_{\rm eff}=P_0VSVS...SVP_0$ obtained at arbitrarily high order in perturbation theory. This implies that all diagonal matrix elements of $\tilde{H}_\mathrm{eff}$ must be identical to any order in perturbation theory, $\braket{\Uparrow_t \Uparrow_b | \tilde{H}_\mathrm{eff}|\Uparrow_t \Uparrow_b} = \braket{\Uparrow_t \Downarrow_b | \tilde{H}_\mathrm{eff}|\Uparrow_t \Downarrow_b} = \dots$, and similarly all off-diagonal matrix elements must be identical (and real), $\braket{\Uparrow_t \Uparrow_b | \tilde{H}_\mathrm{eff}|\Downarrow_t \Downarrow_b}=\braket{\Uparrow_t \Downarrow_b | \tilde{H}_\mathrm{eff}|\Downarrow_t \Uparrow_b}$.
Crucially, this implies that $\tilde{H}_\mathrm{eff}$ becomes trivial if these off-diagonal matrix elements vanish. These off-diagonal elements only emerge at order approx. $L = \sqrt{N}$ (length of a chain) in perturbation theory in $K/J$, since tunneling $\ket{\Uparrow} \to \ket{\Downarrow}$ requires flipping all spins of a given chain, and $H_\mathrm{eff}^{(n)}$ consists of \emph{local} interactions.
This implies that $\braket{\Uparrow_t \Uparrow_b | \tilde{H}_\mathrm{eff}|\Downarrow_t \Downarrow_b} \sim (K/|J|)^L \Delta^{-L} \sim (K/|J|)^L e^{-L}$, where $\Delta >0$ is characteristic dimensionless energy difference between the ground state and excited states. Importantly, this implies that such off-diagonal matrix elements are exponentially supressed with the length of the chains, and in the thermodynamic limit $L\to \infty$, these chains are effectively uncoupled.
We therefore conclude that the ground state has a sub-extensive degeneracy $\mathcal{O}(\sqrt{N})$, consisting of $\sim 2^{\sqrt{N}}$ states corresponding to a two-fold degree of freedom per chain.
Note that our arguments are only valid in the perturbative limit and will eventually break down for $K/J \nless 1$.
Similar states are also obtained in bilayer Kitaev model with Heisenberg interaction for different stacking orders and can be referred to as ``classical'' spin liquids \cite{Seifert_PRb2018}, formed by Ising ``macrospins'' corresponding to the two-fold degenerate chains.

\subsubsection{Coupled dimers for $\mathbf{n} = \hat{z}$}
For $\mathbf{n}=[0,0,1]$, the first order contribution for both $x$ and $y$ bonds vanish. We obtain the following effective Hamiltonian,
\begin{align}
    H_{\rm eff}^{ (1)} &= K \sum_{\langle ij\rangle^z}\eta_i^z\eta_j^z \hspace{20 pt} ~~~~~~ (z \ \rm bonds) \label{eq:13}\\
     H_{\rm eff}^{ (2)} &= \frac{2K^2}{J}\sum_{\langle ij\rangle^{x/y}} \eta_i^x\eta_j^x \hspace{20 pt} ( x/y \ \rm bonds)\label{eq:14}
\end{align}
Note that there are no second or higher order contribution on the $z$-bonds in this case since $[H_K,P_0]=0$, which implies that the higher order contributions in the perturbation theory vanish as $(1-P_0)H_KP_0$ type terms are identically zero. The $H_{\rm eff}^{(1)}$ forms Ising dimers (see Fig.~\ref{fig4}(b)) such that the spins along $z$-bonds are `locked' along the $\hat{z}$-axis, which forms a doublet. A single $H_{\rm eff}^{(2)}$ bond acting on these dimers flips two spins, thereby breaking the Ising dimers. The doublet operators can be expressed as a pseudospin in terms of the $\eta$ degrees of freedom,
\begin{align}
    \rho^z_{in}=& \frac{1}{2}(\eta_{i}^z+\eta_{n}^z )\nonumber \\
    \rho^\pm_{in}=& \eta^\pm_i\eta^\pm_n
\end{align}
where $\rho^z$ is a dipolar and $\rho^x$ and $\rho^y$ are octupolar operators. In terms of the new degrees of freedom, the ground state of Eq.~\eqref{eq:13} are given by the eigenstates of $\rho_{in}^z$. In order to determine if the dimers are coupled via higher order processes, we treat $H_{\rm eff}^{(2)}$  on the $x/y$ bonds as a perturbation on the ground state. We obtain a non-zero contribution involving all four $x/y$ bonds which can be expresses as a ring exchange term.
\begin{align}
    H^{\rm ring}_{\rm eff}=&P_0H_{g2}SH_{g2}SH_{g2}SH_{g2}P_0\nonumber\\
    =&\frac{2K^5}{J^4}\sum_{\hexagon}P_0(\eta^x_i\eta^x_k\eta^x_l\eta^x_n)P_0 \label{eq:16}
\end{align}
where $P_0=\prod_{\langle ij\rangle_z}(1+\eta^z_i\eta_j^z)/2$ and the sum over all the hexagons. In terms of the new pseudospin degrees of freedom, Eq.~\eqref{eq:16} can be expressed as
\begin{align}
   H^{\rm ring}_{\rm eff}=\frac{2K^5}{J^4}\sum_{\hexagon}[\rho^x_{in}\rho^x_{kl}]
   \label{eq:Ising z}
\end{align}
where $\langle in\rangle$ and $\langle kl\rangle$ are the two $z$-bonds belonging to the ring. The ring exchange term couples the dimer degrees of freedom along the $x$ direction and once again forms chains for the octupolar degrees of freedom, $\rho^x$. We also conducted an exact diagonalization study on a 16-site lattice, which included a central hexagonal region, along with two additional $z$-bond connections (see Fig~\ref{fig4}) which agrees with the splitting due to Eq.~\eqref{eq:Ising z} and indicates no further splitting. 

Similar to the previous subsection, here we argue that the chains remain decoupled within the perturbation theory. Considering two adjacent dimer chains, a $\pi$-rotation about the $x$-axis, $U'=\exp{(-i\frac{\pi}{2}}\sum_{\nu, \langle in\rangle}(\sigma^x_{\nu i}+\sigma^x_{\nu n}))$, the dimers along that chain flip from $\ket{\Uparrow}\rightarrow\ket{\Downarrow}$ and vice-versa. Via this rotation, it is possible to map all diagonal matrix elements. The off-diagonal matrix elements require flipping all the spins on the dimer chains, leading to a vanishingly small matrix element in the thermodynamic limit. 

\section{Conclusions} \label{sec:conclusion}
In conclusion, the investigation of the phase diagram of a bilayer Kitaev honeycomb model with Ising interlayer interactions using both perturbative arguments as well as Majorana mean-field theory has yielded valuable insights into the complex interplay between topological order and magnetic tendencies in quantum spin liquids.

When the Kitaev interaction is of the same sign in both layers, we observe a first-order transition from the Kitaev spin liquid state to a magnetically ordered state.

However, when the layers have opposite signs of the Kitaev interaction, our study uncovered a higher stability of the Kitaev spin liquid.
We also find that on a mean-field level, an additional intermediate gapped $\mathbb{Z}_2$ spin liquid state emerges, which ultimately becomes unstable for larger $J/|K|$, when visons are expected to condense and topological order is destroyed. The stability and nature (in particular, topological order) of this intermediate spin liquid is an interesting direction for further study, e.g. using advanced numerical methods. 
The confined phase at large $J/|K| \gg 1$ is aptly described by a highly frustrated $120^\circ$ compass model.

Furthermore, we have performed perturbative analyses for the cases where the Ising axis lies along the $\hat{z}$-axis or in the $xy$ plane. Remarkably, in both instances, we find that 1D Ising chains that intriguingly remain decoupled within perturbation theory, and can be viewed as ``macrospin'' degrees of freedom.
Interesting directions for future studies include exploring different stacking orders, and twisting the two layers, likely leading to a rich interplay of various orders preferred by spatially modulating stacking patterns.

\section{Acknowledgements}
We thank Johannes Knolle and Emilian Nica for fruitful discussions. AV and OE acknowledge support from NSF Award No. DMR 2234352. UFPS was supported by the Deutsche Forschungsgemeinschaft (DFG, German Research Foundation) through a Walter Benjamin fellowship, Project ID 449890867, and the DOE office of BES, through award number DE-SC0020305. This research was supported by the National Science Foundation under Grant No. NSF PHY-1748958.
\\

\appendix

\section{Exact diagonalization for $\textbf{n}=(1,1,0)/\sqrt{2}$} \label{sec:exact-diag}
\begin{figure}
    \centering
    \includegraphics[width=1\linewidth]{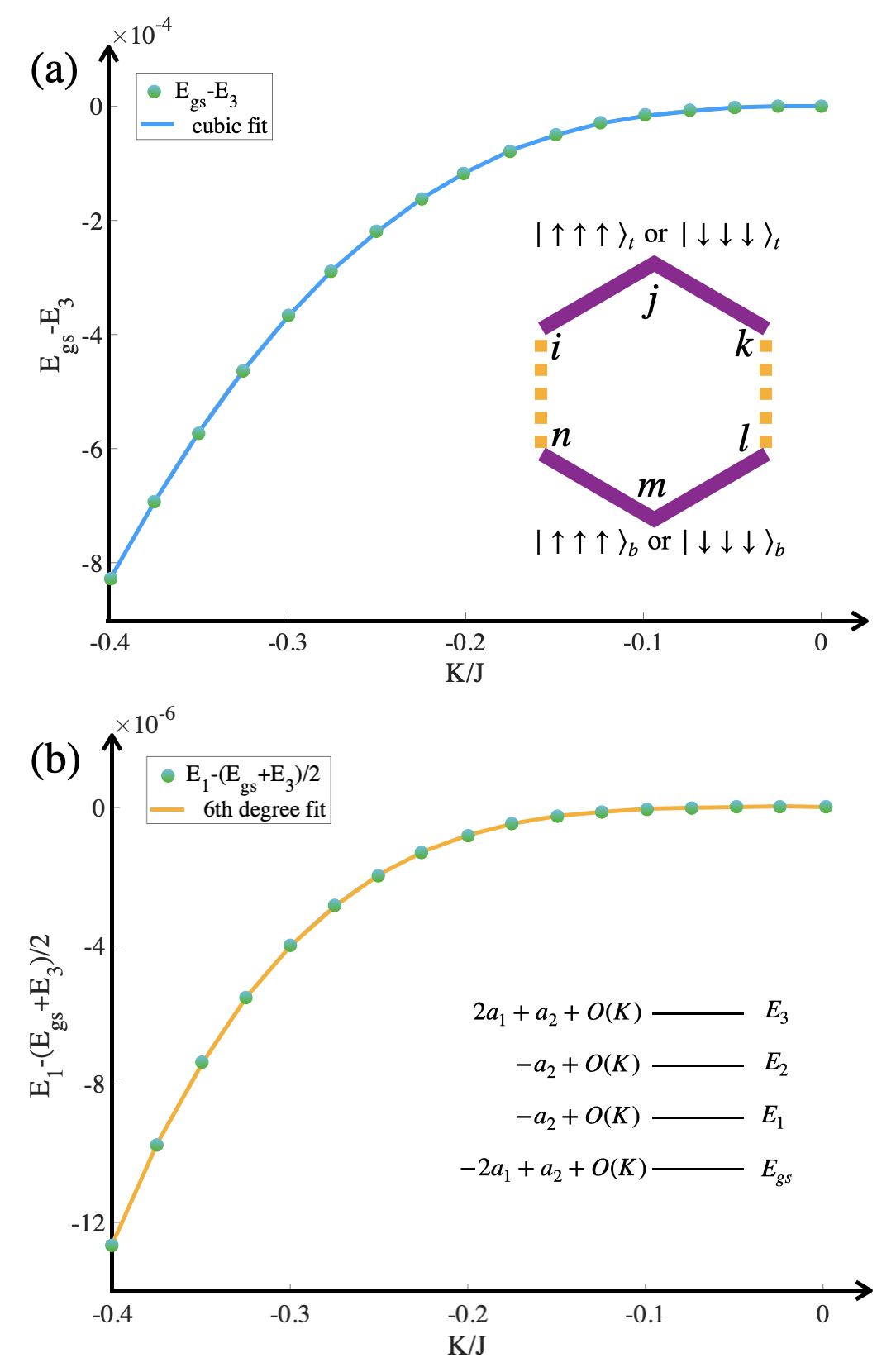}
    \caption{(a), (b) Fitting and relevant energy differences are plotted as a function of $K/J$. Inset (a): Primary hexagon is denoted with the possible unperturbed ``chain" states for top and botton two $x/y$ bonds. (b) Lowest 4 eignevalues, from which coefficient data is extracted.}
    \label{fig:A1}
\end{figure}

We describe here the exact diagonalization calculation of the effective Hamiltonian when $\textbf{n}=(1,1,0)/\sqrt{2}$. Considering a hexagon (12 sites), there are four $x/y$ bonds. The effective Hamiltonian Eq. (\ref{eq:10}) fixes the spins along these bonds to be either $\ket{\uparrow}$ or $\ket{\downarrow}$ state (along the $z$-axis).
The ground state manifold spans: $\ket{\uparrow\uparrow\uparrow}_{t}\ket{\uparrow\uparrow\uparrow}_{b},\ket{\uparrow\uparrow\uparrow}_{t}\ket{\downarrow\downarrow\downarrow}_{b},\ket{\downarrow\downarrow\downarrow}_{t}\ket{\uparrow\uparrow\uparrow}_{b},\ket{\downarrow\downarrow\downarrow}_{t}\ket{\downarrow\downarrow\downarrow}_{b}$, where $t/b$ represent the `top' and `bottom' three spins, see also Fig.~ \ref{fig:A1}. 

To find the coupling between these two segments of the Ising chains, we perform an exact diagonalization on the full Hamiltonian, Eq.~\eqref{eq:K} and Eq.~\eqref{eq:J} for $\mathbf{n}=(1,1,0)/\sqrt{2}$.
The four lowest eigenvalues, and the corresponding eigenvectors are extracted.
In this $4$-dimensional subspace, we perform a rotation of basis to the ground-space basis of Eq.~\eqref{eq:10}, mentioned above.
This 4-dimensional Hamiltonian can be written in terms of spin matrices (up to additional constants): $\Sigma_t^\alpha$ and $\Sigma_b^\alpha$, where $\alpha=x,\ y,\ z$.
\begin{equation}
    H^{\rm ED}_{\hexagon}= a_1(\Sigma^x_t+\Sigma^x_b)+a_2\Sigma^x_t\Sigma^x_b, \label{eq:A1}
\end{equation}
where $a_1$ and $a_2$ are coefficients that we determine in the following steps. First, the eigenvalues, of the above Hamiltonian can be written down as: $E_{gs}=-2a_1+a_2,\ E_{e1}= -a_2,\ E_{e2}=-a_2,\ E_{e3}= 2a_1 +a_2$. In addition, there is an $O(K)$ term in all of these eigenvalues, from the unperturbed Hamiltonian. To extract coefficient $a_1$, eigenvalues $E_{gs}$ and $E_3$ are subtracted, and plotted as a function of $K/J$, Fig.~\ref{fig:A1}(a). A cubic fit suggests that $a_1\approx 0.01 \frac{K^3}{J^2}$. Similarly, for $a_2$, the combination $E_1-(E_{gs}+E_3)/2$ gets rid of the $O(K)$ term and retains $a_2$. Plotting this as a function of $K/J$ and fitting suggests a $6^{th}$ order fit with $a_2\approx 10^{-5} \frac{K^6}{J^5}$.

These operators with their coefficients can be rewritten in terms of the $\eta$ spins, as $\Sigma^x_t=\eta^x_i\eta^x_j\eta^x_k$ and $\Sigma^x_b=\eta^x_l\eta^x_m\eta^x_n$ to obtain Eq. (\ref{HED}). 
\bibliography{references.bib}
\end{document}